\documentclass[a4paper,twocolumn]{esapub}
\usepackage{graphicx}
\usepackage{amssymb}
\usepackage{latexsym}
\usepackage{longtable,lscape,graphicx}
\usepackage{amsmath}
\title{Is the solar corona nonmodally self-heated?}

\author{Bidzina M. Shergelashvili$^{(1),(2)}$, Andria D. Rogava$^{(2),(3)}$ and
Stefaan Poedts$^{(1)}$}

\affil{$^{(1)}$CPA, Katholieke Universiteit Leuven,
Celestijnenlaan 200B, B-3001 Leuven, Belgium
\\

$^{(2)}$CPA, Abastumani Astrophysical
Observatory, Kazbegi ave. 2a, 380060 Tbilisi, Georgia\\

$^{(3)}$Abdus Salam International Centre for Theoretical Physics,
Trieste 34014, Italy}

\begin{document}
\maketitle
\begin{abstract}

Recently it was pointed out that nonmodally (transiently and/or
adiabatically) pre-amplified waves in shear flows, undergoing
subsequent viscous damping, can ultimately heat the ambient flow.
The key ingredient of this process is the ability of waves to
grow, by extracting energy from the spatially inhomogeneous mean
flow. In this paper we examine this mechanism in the context of
the solar coronal plasma flows. ``Self-heating" (SH) processes are
examined when both viscous damping and magnetic resistivity are at
work. We show that if the plasma viscosity is in the favorable
range of values the asymptotic SH rate in these flows can be quite
substantial.

\end{abstract}

\section{Introduction}

Astrophysical plasma flows are often complex, inhomogeneous,
dynamic systems hosting different kinds of collective phenomena.
Sometimes interactions between collective phenomena and ambient
flows feature a highly sophisticated level of complexity. In
general, shear flows (SF) are well-known examples of behavioral
complexity - both in neutral and charged fluids, both for
terrestrial and astrophysical cases and both on the level of
experiments (observations) and theory (simulations) - SF display a
wide spectrum of ``shear-induced" phenomena. In particular,
astrophysical SF are gradually becoming a popular subject of
research because it becomes increasingly more clear that most of
the cosmic plasmas are flowing with spatially inhomogeneous rates
and in most of these flows shear-induced processes may leave a
considerable imprint on the observational appearance of the
related astronomical objects.

A separate class of shear-induced phenomena is related with the
non-self-adjointness of the linear operators governing the
dynamics of SF systems (Trefethen et al. 1993). Fluctuations in a
``parent" SF obey ``non-Hermitian" equations and, therefore, their
evolution, their interaction with the ambient flows can not be
molded within the standard Hamiltonian formalism. The most
striking novelty is the appearance of new, shear-induced
nonperiodic modes - so-called Kelvin or shear vortices.
Interactions between the modes and between the flow and the modes
bring into the game a {\it nonmodal complexity}: these
``non-normal" modes show a transient increase in amplitude, they
are coupled and are transformed into each other, they may feature
beating phenomena and may become the subjects of different shear
instabilities.

Recently it was pointed out that SF may host yet another
interesting non-modal phenomenon related with the combined
presence of nonmodality and dissipative processes in the flows. It
can formally be considered as a three-phase process (Rogava 2004):

\begin{enumerate}

\item waves and/or vortices get \emph{excited }within a flow;

\item they \emph{amplify} nonmodally, due to the presence of the
shear flow, extracting a part of the flow kinetic energy;

\item high-amplitude waves and/or transiently amplified vortices
undergo a viscous decay and/or magnetic diffusion and give, in the
form of heat, a part of their energy back to the flow.

\end{enumerate}

As a result of this three-step process the fluctuation gives back
to the flow {\it more} thermal energy than it had at the moment of
its excitation. Repeated continuously, throughout the whole volume
of the flow, these elementary processes should lead to a net
heating of the flowing plasma. The process was named
\emph{``self-heating"} (SH), because what actually happens is that
nonmodality manages to transfer a part of the flow's regular
kinetic energy into a thermal energy; i.e., the heating happens
conservatively without any outside source of energy, the flow
manages to heat itself.

The possibility of this process was originally surmised in (Rogava
2003a) and later it was examined and specified in detail for the
relatively simple case of a neutral fluid SF and sound waves
sustained by it (Rogava 2004). It was shown that in the
hydrodynamic situation - even when a mixture of sound waves and
Kelvin vortices is propagating in a simple, plane-parallel
Couette-like SF and, therefore, acoustic waves are able to amplify
only linearly, extracting the kinetic energy of the ambient flow -
the self-heating rate can be quite high: fluctuations are able to
give back to the flow several times more energy than they have
initially had.

The purpose of this paper is to explore whether the SH can be
efficient in magnetized plasma MHD flows. In this particular study
we consider only the case of a plane-parallel flow with a linear
velocity profile. However, in order to lay the foundation for a
further, more general, consideration of flows with kinematic
complexity we derive our equations for the case of swirling
(helical) flows, following the approach developed in (Rogava et
al. 2003a and 2003b) (henceforth referred to as R1 and R2) and
consider the case of the plane-parallel flow as a particular
example.

Our results show that both in the incompressible  (when the flow
sustains only Alfv\'en waves) and in the compressible (full
spectrum of MHD waves) cases the self-heating is even more
efficient than it was in the hydrodynamic case. Our numerical
simulations revealed that the values of the dimensionless
asymptotic SH rate $\Xi_{\infty}$ can be of the order of several
tens. This allows us to argue that the SH may easily be one of the
robust physical mechanisms contributing to the heating of the
solar corona through the viscous and/or resistive damping of
nonmodally pre-amplified waves.

It should be noted that nonmodal SH, being efficient \emph{per se}
possesses another attractive feature in the context of its
possible relevance for coronal heating: all existing wave heating
scenarios face common problem: how is the energy transferred from
longer length-scales to shorter ones, where dissipative effects
are significant?! SH phenomena, being nonmodal, involve the
shear-induced drift of the wave number vector ${\bf k}(t)$, which
provides a natural, flow-related mechanism for the gradual
decrease of the mode's length-scale. This process, being linear in
plane-parallel SF, may have an exponentially fast nature in
geometrically and kinematically more sophisticated cases.
Therefore, we envisage that kinematically complex flow patterns,
such as solar spicules, macrospicules and tornados, magnetic
plumes, might host much more efficient, sometimes even explosive,
SH events.

\section{Main consideration}

The present study follows  the theory of nonmodal phenomena in
helical MHD flows, recently developed both for the incompressible
(R1) and the compressible (R2) cases. The difference between our
current and those studies is in the presence of a dissipation -
viscous damping and/or magnetic resistivity - represented in the
equations by terms proportional to the coefficients of kinematic
viscosity ($\nu_h$) and magnetic resistivity ($\nu_m$),
respectively. Since a nonmodal analysis is essentially linear, the
instantaneous values of all physical variables have to be splitted
into their equilibrium and fluctuative components: ${\bf
B}{\equiv} {\bf B}_0+{\bf B}^{\prime}$,
${\rho}{\equiv}{\rho}_0+{\rho}^{\prime}$, etc; with the subsequent
linearization of the equations for perturbations.

The equilibrium, considered in R1 and R2, assumes a homogeneous
MHD plasma ($\rho_0=const$) flow, embedded in a homogeneous,
vertical magnetic field (${\bf B}_0{\equiv}[0,~0,~ B_0=const]$).
The mean flow vector field ${\bf U}_0(r)$ in the R1 was specified
as:
$$
{\bf U}(r){\equiv}[0,~r{\Omega}(r),~U(r)], \eqno(1)
$$
with ${\Omega}(r)={\cal A}/r^n$, where $r=(x^2+y^2)^{1/2}$ is a
distance from the rotation axis, while ${\cal A}$ and $n$ are some
constants.

If we take into account the viscous damping and magnetic
resistivity then linearized equations for fluctuations can be
written as [${\cal D}_t{\equiv}{\partial}_t+({\bf
U}_0{\cdot}{\nabla})$]:
$$
{\cal D}_td+{\nabla}{\cdot}{\bf u}=0, \eqno(2)
$$
$$
{\cal D}_t{\bf u}+({\bf u}{\cdot}{\nabla}){\bf U}_0=-{\nabla}p+
C_A^2[{\partial}_z{\bf b}-{\nabla}b_z]  + \nu_h \Delta {\bf u},
\eqno(3)
$$
$$
{\cal D}_t{\bf b}=({\bf b}{\cdot}{\nabla}){\bf
U}_0+{\partial}_z{\bf u} +{\bf e}_z({\nabla}{\cdot}{\bf u}) +
\nu_m \Delta {\bf b}, \eqno(4)
$$
$$
{\nabla}{\cdot}{\bf b}=0, \eqno(5)
$$
with $d$ ${\equiv}$ ${\rho}^{\prime}/{\rho}_0$, $p$ ${\equiv}$
$p^{\prime}/{\rho}_0$ and ${\bf b}$ ${\equiv}$ ${\bf B}^{\prime}
/B_0$. Note that for \emph{incompressible fluctuations} instead of
(2) we have:
$$
{\nabla}{\cdot}{\bf u}=0, \eqno(6a)
$$
while for the \emph{compressible} case the closure of the (2-5) is
guaranteed by the equation of state, implying:
$$
p^{\prime} = C_s^2 \rho^{\prime}, \eqno(6b)
$$
with $C_s$ and $C_A$ being the homogeneous speed of sound and the
Alfv\'en speed, respectively.

The nonmodal method for studying the dynamics of linearized,
small-scale fluctuations in kinematically complex flows (Lagnado
et al. 1984, Craik and Criminale 1986, Mahajan and Rogava 1999)
enables the reduction of the initial set of partial differential
equations for the perturbation variables $F({\bf r},t)$ in the
real physical space to initial value problem for the spatial
Fourier harmonics (SFH) of the same variables, ${\hat F}({\bf
k},t)$), defined in the ${\bf k}$-space. The key element of this
approach is the time variability of the ${\bf k}(t)$'s, imposed by
the presence of the shear flow and governed by the following set
of equations [${\partial}_t^nf{\equiv}f^{(n)}$]:
$$
{\bf k}^{(1)}+{\cal S}^T\cdot{\bf k}=0,  \eqno(7)
$$
which gives, depending on the particular form of the shear matrix
{\cal S} (Mahajan and Rogava 1999), a full evolutionary picture of
the temporal drift of the wave number vector field ${\bf k}(t)$.

For a helical flow, specified by the equilibrium velocity (1) and
with five nonzero components of the traceless shear matrix
(specifying the stretching of flow-lines ($\sigma$), velocity
shear in rotational ($A_1$ and $A_2$) and ejectional ($C_1$ and
$C_2$) components of the velocity, respectively) $k_z$ stays
constant, while the transversal components obey (R1):
$$
k_x^{(1)}+{\sigma}k_x+A_2k_y+C_1k_z=0, \eqno(8a)
$$
$$
k_y^{(1)}+A_1k_x-{\sigma}k_y+C_2k_z=0, \eqno(8b)
$$
implying that $k_x(t)$ and $k_y(t)$ may have an algebraic,
exponential or periodic time dependence.

The ordinary nonautonomous differential equations for the SFH of
physical variables can be derived from the set (2--5) and written
in the following way:
$$
{\varrho}^{(1)}={\bf k} \cdot {\bf v}, \eqno(9)
$$
$$
v_x^{(1)}+({\cal S}\cdot {\bf v})_x=-k_x{\cal P} + C_A^2(k_z b_x
-k_x b_z) - {\nu}_h|{\bf k}|^2v_x, \eqno(10a)
$$
$$
v_y^{(1)}+({\cal S}\cdot {\bf v})_y=-k_y{\cal P} + C_A^2(k_z b_y -
k_y b_z) - {\nu}_h|{\bf k}|^2v_y, \eqno(10b)
$$
$$
v_z^{(1)}+({\cal S}\cdot {\bf v})_z=-k_z{\cal P} - {\nu}_h|{\bf
k}|^2v_z, \eqno(10c)
$$
$$
b_x^{(1)}=({\cal S}\cdot {\bf b})_x - v_x-\nu _m |{\bf k}|^2 b_x,
\eqno(11a)
$$
$$
b_y^{(1)}=({\cal S}\cdot {\bf b})_y - v_y-\nu _m |{\bf k}|^2 b_y,
\eqno(11b)
$$
$$
{\bf k} \cdot {\bf b}= 0. \eqno(12)
$$

The total energy of the perturbation consists of a compressional,
a kinetic and a magnetic part (the first part is absent in the
incompressible limit) and is equal to:
$$
E \equiv [C_s^2{\varrho}^2 + C_A^2{\bf b}^2 + {\bf v}^2]/2,
\eqno(13)
$$
This total energy obeys the following nonautonomous equation:
$$
E^{(1)}=(A_1+A_2)(b_xb_y-v_xv_y) +
$$
$$
+C_1(b_xb_z-v_xv_z)+C_2(b_yb_z-v_yv_z) +
$$
$$
+{\sigma}[(b_x^2-b_y^2)-(v_x^2-v_y^2)]-|{\bf k}|^2[\nu_h{\bf v}^2
+ \nu_m {\bf b}^2]. \eqno(14)
$$

Finally, following the hydrodynamic case (Rogava 2004), we define
the asymptotic \emph{self-heating rate} as the limit
$$
\Xi_{\infty} \equiv \lim_{t \to \infty}\Xi(t), \eqno(15)
$$
of the following function:
$$
\Xi(t) \equiv {1 \over{E(0)}} \int_{0}^{t}{\biggl[\nu_h {\bf
v}^2(t')+ \nu_m {\bf b}^2(t')\biggr]}|{\bf k}|^2(t')dt', \eqno(16)
$$

In this short paper our purpose is {\it not} the study of fully
complex helical flows. Rather, we wish to demonstrate the
efficiency of the self-heating for relatively simple,
plane-parallel velocity patterns of magnetized plasmas. Therefore,
we consider only the simplest, plane-parallel flow case both for
incompressible and compressible fluctuations.

\subsection{Incompressible limit}

%%%%%%%%%%%%%%%%%%%%%%%%%%%%%%%%%%%%%%%%%%%%%%%%%%%%%%%%%
\begin{figure*}
\centering
\includegraphics[scale=0.8]{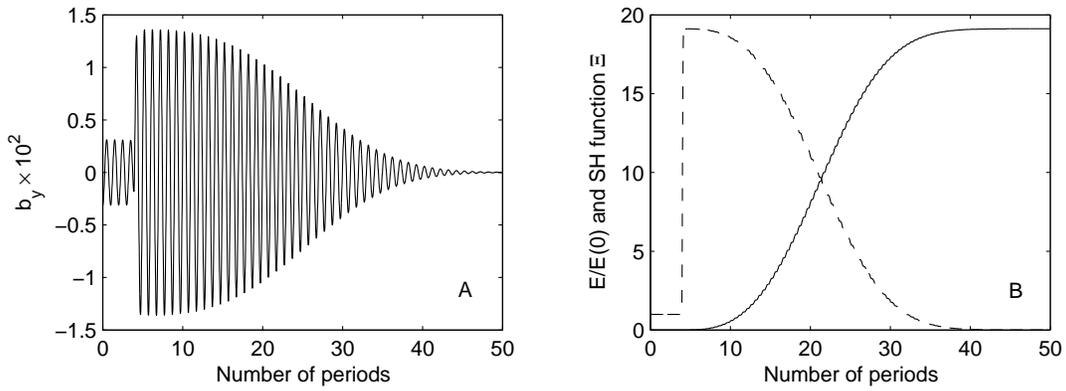}
\caption{The temporal evolution of the $b_y$ component of the
magnetic field perturbation (panel A) and the normalized energy
$E/E(0)$ and self-heating function (panel B) for the following
values of medium and wave parameters: $V_A= 100$ km s$^{-1}$,
$C_x=0.08$ s$^{-1}$, $C_y=0$ s$^{-1}$, $\eta=10^4$ cm$^2$
s$^{-1}$, $\nu = 10^9$ cm$^2$ s$^{-1}$, $k_z=4\cdot 10^{-9}$
cm$^{-1}$ corresponding to the oscillation period $T\approx 2.6 $
min. \label{fig1i}}
\end{figure*}
%%%%%%%%%%%%%%%%%%%%%%%%%%%%%%%%%%%%%%%%%%%%%%%

In this case (6a) holds leading to the algebraic relation: $({\bf
k\cdot v})=0$. The flow sustains only Alfv\'en waves, strongly
modified by the presence of the SF.

The efficiency of the wave heating mechanism strongly depends on
temporal scales of the wave excitation/damping. In our case,
incompressible MHD disturbances are able to amplify transiently,
within a very short time interval, compared to the total timescale
of the wave evolution (Chagelishvili et al. 1993). This makes this
process a quite efficient mechanism for the excitement of
large-amplitude Alfv\'en waves. However, when dissipation (viscous
damping  and/or magnetic resistivity) is also taken into account,
it is reasonable to expect that under favorable conditions the
``pre-amplified" Alfv\'en  waves will eventually give their
energy\footnote{Which they  have just  ``stolen" from the flow via
shear-induced transient amplification!} back to the flow in the
form of heat.

 In order to give an illustrative example we solved the set of
equations numerically for different values of the parameters,
roughly typical for various structures of the solar atmosphere:
$V_A= 100$ km s$^{-1}$, $C_x=0.08$ s$^{-1}$, $C_y=0$ s$^{-1}$ and
a fixed value of wavenumber $k_z=4\cdot 10^{-9}$ cm$^{-1}$
corresponding to the oscillation period $T\approx 2.6 $ min. In
all our calculations we used the conventional value of the
magnetic resistivity coefficient (see Walsh \& Ireland, 2003):
$\nu_m=10^4 cm^2 s^{-1}$. As regards the viscosity coefficient,
since the presence or absence of the microturbulence may change it
by many orders of magnitude (Ruderman et al. 1998), we were
inclined to consider it as a free parameter and we made a
numerical analysis for different values.

It was found that when the shear viscosity is determined only by
the momentum transfer due to ion diffusion and, therefore $\nu_h
\simeq \nu_m$ (Ruderman et al. 1998) the actual timescale of the
viscous damping is too large. Therefore, the damping of these
waves in a laminar flow with low values of $\nu_h$ and $\nu_m$
could hardly account for the spatially confined processes of the
coronal heating.

However, if one assumes that a microturbulence can be present in
the coronal plasma, it may increase the viscosity coefficient in a
very considerable way. In this case the temporal scale of the
viscous damping for transiently pre-amplified waves drastically
changes. In particular, it was found that when  $\nu_h \ge 10^9$
cm$^2$ s$^{-1}$, Alfv\'en waves with the above-specified
parameters damp efficiently within physically reasonable time
intervals. In other words, the viscosity coefficient has to be
increased by at least a factor of $10^5$, compared to its laminar
value, in order to guarantee effective damping. Still, this is
about five orders of magnitude \emph{less} than values of the
$\nu_h$ necessary for the efficient coronal heating via the
conventional wave heating mechanisms (Ruderman et al. 1998). In
the case of the SH of transiently pre-amplified Alfv\'en waves
efficient dissipation occurs even for these, relatively ``mild"
($\sim 10^9$cm$^2$ s$^{-1}$) values of the viscosity coefficient!

Since the waves were considerably amplified before starting to
decay it is reasonable to expect that they give back to the flow
\emph{much more} energy than they initially had at the moment of
excitation. This means that the self-heating mechanism works and
the plasma gets heated via the combined action of nonmodal
transient pre-amplification of waves and their subsequent viscous
decay. In panel~A of Fig.~\ref{fig1i} we show the temporal
evolution of the magnetic field component $b_y$ (the values are
scaled up by the factor $10^2$). In panel~B of the same figure the
curves showing the temporal behaviour of the normalized total
energy $E/E(0)$ of the perturbation (dashed line) and the SH rate
function $\Xi(t)$ (solid line) are presented.

%%%%%%%%%%%%%%%%%%%%%%%%%%%%%%%%
\subsection{The compressible medium}
%%%%%%%%%%%%%%%%%%%%%%%%%%%%%%%%

For compressible perturbations the closure of the set of equations
comes from the Eq. (6b). In this case all three MHD waves can
exist and in shear flows their nonmodal mutual transformation
(Chagelishvili, Rogava and Tsiklauri 1996) may take place. Since
we are interested in the perspectives for the nonmodal SH in solar
coronal flows, we have to concentrate on the low plasma $\beta$
case. In this case the slow mode is decoupled from two other MHD
wave modes, but the Alfv\'en and fast magnetosonic modes are
coupled and may transform into each other (Rogava, Poedts and
Mahajan 2000).

\begin{figure}
\centering
\includegraphics[scale=0.8]{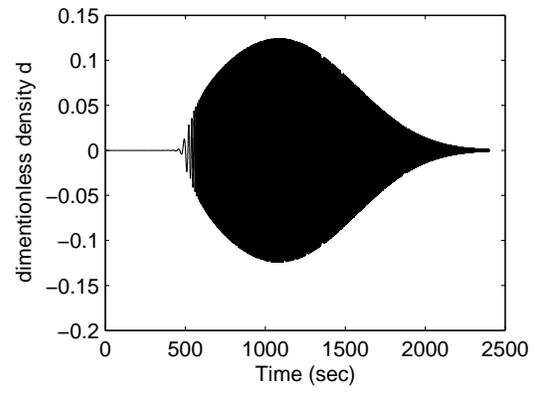}
\caption{The dimensionless density perturbation $d$ vs. time (seconds) for the values of
the
medium and wave parameters: $C_s=5\cdot 10^6$ cm s$^{-1}$, $V_A=8\cdot 10^7$ cm s$^{-1}$,
$\eta=10^{4}$ cm$^2$ s$^{-1}$, $\nu =5\cdot 10^{11}$ cm$^2$ s$^{-1}$, $C_x=0.04$ s$^{-1}$
and $C_y=0$ s$^{-1}$.\label{fig1c}}
\end{figure}

\begin{figure}
\centering
\includegraphics[scale=0.45]{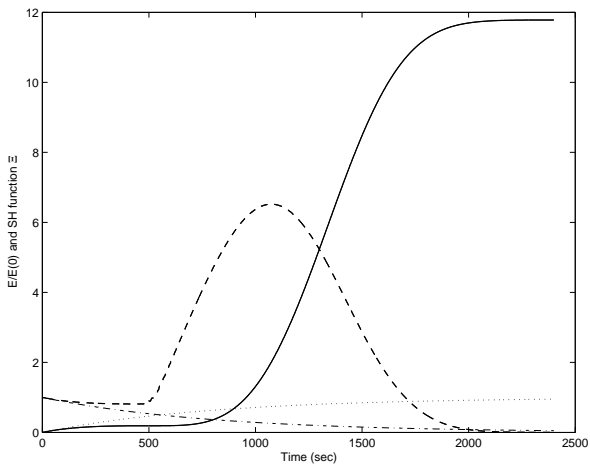}
\caption{The temporal evolution of the normalized energy $E/E(0)$
(dashed line) and self-heating function (solid line) corresponding
to the solution given in Fig.~\ref{fig1c}. By the dash-dotted and
dotted lines the same quantities are shown, respectively, for the
case of no shear flow $C_x=0$. \label{fig2c}}
\end{figure}

The governing equations were solved numerically for the following
set of parameters: $C_s=5\cdot 10^6$ cm s$^{-1}$, $V_A=8\cdot
10^7$ cm s$^{-1}$, $\eta=10^{4}$ cm$^2$ s$^{-1}$, and $C_x=0.04$
s$^{-1}$. In Fig.~\ref{fig1c} the temporal evolution of the
dimensionless density perturbation is plotted. In addition, we
present in Fig~\ref{fig2c} the curves of variation of the
normalized total energy $E/E(0)$ of perturbations and the function
$\Xi(t)$ of the SH rate.

The evolution of the initially excited Alfv\'en wave formally may
be splitted into three stages:

%\vskip 0.1cm
1) In the initial stage $t\lesssim 500$ s ($k_x(t)>0$) we have a
pure Alfv\'en mode and its energy slowly decreases in time (thick
dashed line in Fig.\ref{fig2c}). Here we also show the evolution
of the total energy in the shearless limit ($C_x=0$) by the thin
dashed-dotted line. We can see that at this stage of the
evolution, in both cases, the energy evolves similarly: these
curves almost coincide with each other. The slow dissipation of
the mode is represented by the slight increase of the SH function
both in the case of the non-uniform flow (thick solid line) and in
the case of the uniform (shearless)  flow (thin dotted line).

%%%%%%%%%%%%%%%%%%%%%%%%%%%%%%%%%%%%%%%%%%%%%%%
%%%%%%%%%%%%%%%%%%%%%%%%%%%%%%%%%%%%%%%%%%%%%%%

2) At a certain stage (around the moment of time when $k_x(t)
\approx 0$) the Alfv\'en wave partially transforms into the fast
mode. This fact is represented by the excitation of the density
perturbation, see Fig.~\ref{fig1c}. The perturbation starts
extracting energy from the mean flow\footnote{This situation
drastically differs from the homogeneous flow case, where the
transformation of the Alfv\'en wave into the fast mode does
\emph{not} happen and the energy of the perturbation continuously
decreases (dashed dot line in Fig.~\ref{fig2c})} (dashed line in
Fig.~\ref{fig2c}). The presence of the velocity shear leads to the
further monotonous decrease of the wavelength of the excited fast
mode, while $k_x (t)$ changes the sign and $|{\bf k}|$ starts
growing again, and the energy of the fast mode grows
adiabatically. But the efficiency of the dissipation grows as
well, as far as the length-scales become shorter, and at a certain
stage the energy of the fluctuation ceases to increase and starts
decreasing - the exponential damping becomes prevalent, it starts
dominating over the linear, adiabatic increase of the wave energy.

%%%%%%%%%%%%%%%%%%%%%%%%%%%%%%%%%%%%%%%%%%%%%%%%%%%%%%%%%
%%%%%%%%%%%%%%%%%%%%%%%%%%%%%%%%%%%%%%%%%%%%%%%

3) At the final stage of the evolution the dissipation effects
prevail and the mode starts damping efficiently (see
Fig.~\ref{fig1c}). Eventually all the energy extracted on the
earlier stage by the wave from the flow returns back to the flow
in the form of thermal energy (heat). The SH function rapidly
grows and reaches its asymptotic value (see plateau in Fig.
~\ref{fig2c}) when the wave is completely dissipated. This value
is approximately equal to the doubled maximum value of the
normalized wave energy. This situation is different from the
previously considered incompressible case, where the amplification
of the perturbation had a transient nature and occurred only
within a very small time interval. In that case the quantity of
the produced heat was roughly equal to the wave energy maximum
(see panel B of Fig.~\ref{fig1i}).

In addition, we studied numerically the temporal evolution of the
Alfv\'en disturbances for three different initial values of $k_z$
for fixed values of all other parameters. The results show that
with the increase of the initial wave number, for given values of
dissipation coefficients, the SH rate decreases - the shorter the
initial wavelengths are, the sooner the time comes when the
damping effects prevail. Therefore, the nonmodally excited fast
mode has less time to gain energy from the flow. A more efficient
SH in this case would require smaller values of the dissipation
coefficient. On the other hand, this process has another limit: if
we decrease the value of the initial wavenumber the mentioned
critical moment of time comes too late and, therefore, the
effective time scale of the wave damping increases too much: the
timescale of the heat production becomes much larger than the
characteristic timescales related to the coronal situation.

Similarly, we have different regimes when we study the
self-heating process for a fixed value of  the wavenumber, but for
different values of the viscosity coefficient. For a given
Alfv\'en mode there exist values of the viscosity coefficient for
which the wave amplification process is very intensive but the
effective time of transferring this energy back to the background
medium is too large to be interesting in the solar coronal
context. The SH process becomes most efficient for larger
viscosities. However, further increase of the strength of the
dissipation, again, decreases the SH rate, because the excited
waves are damped too rapidly, without significant nonmodal
amplification by the shear flow. Therefore, there is a limited
range of favorable viscosity levels, at which the self-heating
might be expected to be most efficient.

\section{Conclusions}

The purpose of the present study was to clarify whether the
shear-induced amplification of MHD waves coupled with viscous and
magnetic-resistive  dissipation may lead to a significant
\emph{self-heating} of the SF. In particular, our aim was to see
whether this mechanism could be efficient  for solar coronal
plasma flows and could, arguably, contribute to the heating of the
solar corona.

Since the variety of solar plasma flows is quite wide, both in
terms of geometry and kinematics, the \emph{local} formalism
employed and developed within this study is apt for an arbitrary
velocity pattern with the condition that the involved MHD waves
are having length-scales, ${\l }$, sufficiently smaller than the
linear length-scale,  ${\cal L}$, of a ``parent" flow pattern.
Basically, our consideration follows Rogava et al. (2003a, 2003b)
with addition of dissipative effects related to the presence of
viscosity and magnetic resistivity.

In this paper we have examined only the case of the simplest
velocity pattern with plane-parallel and linear velocity profile.
Both incompressible and compressible limits were investigated. It
was found that:

\begin{enumerate}

\item Incompressible, Alfv\'enic, perturbations, in the ideal MHD
limit, are known to undergo an algebraic instability (``transient
increase") (Chagelishvili et al. 1993; Rogava et al. 1996, Rogava
et al. 2003a), which can be quite strong: under favourable
conditions the total energy of a fluctuation may increase several
hundred times. We found that when dissipation is taken into
account these high-amplitude pre-amplified Alfv\'en waves get
damped and give their energy back to the ambient flow. The
resulting SH is quite substantial. As it could be expected the
asymptotic self-heating rate in this case is quite large: it
increases with the decrease of the Alfv\'en velocity and in the
cases considered here could reach a value of the order of several
tens. This happened to be a case for waves with the characteristic
periods of the order of few minutes.

\item In the compressible case we considered the evolution of an
initially excited Alfv\'en wave (as an example we considered
Alfv\'en mode with period of the order of 30 seconds) in a
plane-parallel flow pattern with the value of the plasma-$\beta$
of the order of $10^{-2}$, typical for the coronal environment. In
this case Alfv\'en waves  and fast magnetosonic waves are coupled,
which leads to the transformation of initially Alfv\'enic
fluctuations into the fast ones \footnote{The latter wave, being
of the acoustic nature, dissipates quite similarly to plain sound
waves, recently shown to be instrumental in the SH of compressible
fluid shear flows (Rogava 2004).} (Rogava et al. 2000). We found
(see Fig.~\ref{fig2c}) that this process, coupled with the
presence of viscous and resistive dissipation, ensures the SH of
the ``parent" flow with the asymptotic rate of the order of
several tens.

\end{enumerate}

These results are quite encouraging. We see that even the simplest
kinds of magnetized plasma SF's are able to heat themselves via
the agency of nonmodally pre-amplified Alfv\'en waves. It is both
tempting and reasonable to surmise that in flow patterns with a
higher degree of kinematic complexity, where nonmodal processes of
energy exchange between flows and waves have a more intense nature
(Mahajan and Rogava, 1999), the resulting SH can be considerably
higher! In certain cases \footnote{For instance, in flows with a
helical mode of plasma motion (Rogava et al. 2003a, 2003b)}
individual flow patterns may tend to self-destruction, undergoing
self-imposed, catastrophically fast SH of an eruptive, explosive
nature. Further studies in this vein are currently being carried
out and the results will be reported elsewhere.

The astrophysically relevant conclusion of the present study is
that nonmodal self-heating, even for relatively simple kinds of
flows, parallel to the magnetic field and with shear profiles
across the field, can pay a significant contribution in the (still
poorly understood) phenomenon of solar coronal heating.

\section*{Acknowledgments}

A.R. is grateful to the CPA/K.U. Leuven for the hospitality in the
period of July-September, 2004. This work has been developed in
the framework of the pre-doctoral program of B.M.S. at the CPA,
K.U.Leuven (scholarship OE/04/33).
%%%%%%%%%%%%%%%%%%%%%%%%%%%%%%%%%%%%%%%%%

\end{document}